\documentstyle[aps,prl,twocolumn,graphicx]{revtex}
\begin{document}
\wideabs{	
\title{ Anomalous NMR response of quasicrystalline icosahedral 
Al$_{72.4}$Pd$_{20.5}$Mn$_{7.1}$ at low temperatures}

\author{J. L. Gavilano,$^{1}$ D. Rau,$^{1}$ Sh. Mushkolaj,$^{1}$   H. 
R. Ott,$^{1}$ J. Dolin${\rm \check s}$ek,$^{2}$ and K. Urban$^{3}$}

\address {
$^{1}$Laboratorium f\"{u}r Festk\"{o}rperphysik, ETH-H\"{o}nggerberg, 
CH-8093 Z\"{u}rich, Switzerland \\
$^{2}$J. Stefan Institute, University of Ljubljana, Jamova 
39, SLO-1000 Ljubljana, Slovenia\\
$^{3}$Institut f\"ur Festk\"orperforschung, Forschungzentrum J\"ulich 
GmbH, D-52425 J\"ulich, Germany} 

\maketitle

\begin{abstract}
 We report the observation of an anomalous $^{27}$Al-NMR response of a
 single grain Al$_{72.4}$Pd$_{20.5}$Mn$_{7.1}$ icosahedral
 quasicrystal at low temperatures.  In an external magnetic field of
 6 T and upon decreasing temperature, we observe a sharp 100 \%
 increase of the resonance linewidth at 2.5 K. No further changes of
 the linewidth are observed down to 0.05 K. The linewidth enhancement
 is accompanied by a small but distinct increase of the spin-lattice
 relaxation rate $T_{1}^{-1}$ and by a maximum 
 of the spin-spin relaxation time $T_{2}(T)$.  All these
 anomalies are absent in external fields of 2.5 T and below.  Our
 observations indicate unusual variations in the stability of isolated 
 magnetic moments in a quasiperiodic metallic environment.
\end{abstract}
\vspace{-0.9 cm}
}    
Various physical properties of icosahedral quasicrystals of Al-Pd-Mn
alloys have been studied in recent years.  Especially intriguing are
their magnetic properties\cite{Chernikov93,Simonet98,Dolinsek01}.  For
example, from data of the magnetic susceptibility and the specific
heat, as well as from the results of calculations of the electronic
structure,\cite{Chernikov93,Krajci98,deLaissardiere00} it has been
inferred that in these materials only a small fraction, of the order
of 1\%, of Mn ions carry a magnetic moment.  The coexistence of
magnetic and nonmagnetic Mn sites in Al-rich quasicrystals has
also been discussed for Al-Pd-Mn quasicrystals with slightly
varying chemical composition.\cite{Bennett87,Warren86,Shinohara93}. 
The formation or the absence of ionic magnetic moments at the Mn sites
in Al-Pd-Mn quasicrystals has first been attributed to differences in
the local chemical environment of the Mn ions.  In particular it was
claimed\cite{Krajci98} that a weak Al-$p$-Mn-$d$ hybridization leads
to the formation of well localized and rather large Mn moments.  More
recently, however, it has been suggested\cite{deLaissardiere00} that
the local environment of the Mn ions alone cannot explain why only few
Mn ions carry a magnetic moment and that also Mn-Mn interactions over
large distances ought to be taken into account.

Previous $^{27}$Al NMR experiments\cite{Dolinsek01} on a single grain
quasicrystal with a composition of Al$_{72.4}$Pd$_{20.5}$Mn$_{7.1}$
have shown that the small number of magnetic Mn ions decreases even
further with decreasing temperature below approximately 20 K. This
unusual behavior in this temperature range was revealed by a reduction
of the NMR linewidth with decreasing temperatures, as well as by
indicative features of the temperature dependence of the magnetic
susceptibility $\chi(T)$.  So far, the cause for these observations is
not clear.

In this work we present experimental evidence for additional anomalies
in the temperature dependences of the NMR spectra and the spin lattice
relaxation rate $T_{1}^{-1}$ of an icosahedral
Al$_{72.4}$Pd$_{20.5}$Mn$_{7.1}$ quasicrystal.  In external magnetic
fields of the order of 6 T and with decreasing temperature we observe
drastic changes in the temperature dependences of both the $^{27}$Al-
and the $^{55}$Mn-NMR linewidth at $T_{b} = 2.5$ K, accompanied by
distinct variations in the temperature dependences of both the
spin-lattice and the spin-spin relaxation rates $T_{1}^{-1}(T)$ and
$T_{2}^{-1}(T)$.  None of these anomalies is observed in fields of the
order of 2.5 T or smaller.

The investigated sample was a single-grain piece of quasicrystalline
material with a nominal composition of
Al$_{72.4}$Pd$_{20.5}$Mn$_{7.1}$, grown by the Czochralski technique
and annealed at 850 $^{\circ}$C. Its chemical and structural quality
has been discussed in ref.  9.  The same sample, previously used to
perform measurements of $^{27}$Al-NMR spectra at higher
temperatures\cite{Dolinsek01}, has also been characterized by
measurements of the magnetic susceptibility and the electrical
conductivity between 1.5 and 300 K. In all our NMR experiments,
standard spin-echo techniques have been employed.
\begin{figure}[t]
\begin{center}\leavevmode
\includegraphics[width=0.65\linewidth]{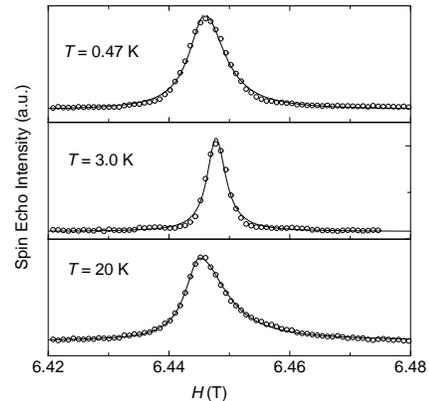}
\caption{ 
Central part of the $^{27}$Al-NMR spectra of 
Al$_{72.4}$Pd$_{20.5}$Mn$_{7.1}$ icosahedral quasicrystal measured at 
71.33 MHz and three temperatures.  The solid lines represent the
best fits to the data using a Lorentzian-type function with a degree
of asymmetry built into it. 
}
\protect 
\label{Figure1}
\end{center}
\end{figure}
  In Fig.  1 we present the central part of the $^{27}$Al-NMR spectra
  of our sample, measured at the Larmor frequency of 71.33 MHz, at
  three different temperatures.  Care has been taken to ensure that
  the relevant experimental parameters are identical for each series
  of measurements.  The well defined peaks in the spectra represent
  the central Zeeman transition ($1/2 \leftrightarrow -1/2$) of the Al
  nuclei.  The wings of the spectrum, $i.e.$, the $\pm 1/2
  \leftrightarrow \pm 3/2$ and $\pm 3/2 \leftrightarrow \pm 5/2$
  transitions caused by the electric quadrupolar perturbation of the
  Al nuclear Zeeman levels ($I = 5/2$) are distributed over a broad
  range of resonant fields, a feature that seems to be generic for
  quasicrystals\cite{Shastry94}.  The full NMR spectrum of our
  material is shown in ref.  3, revealing that the temperature
  dependence of the wings does not exhibit any anomalies at low
  temperatures.  Thus the sudden increase of the NMR linewidth
  observed at 2.5 K and shown in Fig.  2 is of magnetic, and not of
  electric quadrupolar origin.  Hence, the following analysis and
  discussion consider the central transition.
  
\begin{figure}[t]
\begin{center}\leavevmode
\includegraphics[width=0.65\linewidth]{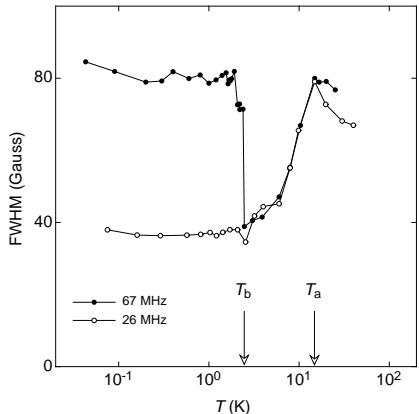}
\caption{ $^{27}$Al-NMR linewidth $\Delta$ as a function of
temperature for data measured at 26.00 and 67.00 MHz.  Sharp features
are observed in $\Delta(T)$ for the data measured at 67 MHz at $T_{a}
= 15$ K and $T_{b} = 2.5 $ K. The solid lines are guides to the eye }
\protect
\label{Figure2}
\end{center}
\end{figure}
 Two sharp features can clearly be distinguished in the temperature
 dependence of the $^{27}$Al-NMR linewidth $\Delta$ ($FWHM$) measured
 at 67 MHz (see Fig.  2).  These are, a discontinuity in $\Delta$ and
 a break in its slope at $T_{b} = 2.5$ K and $T_{a} \approx 15$ K,
 respectively.  As may also be seen in the same figure, $\Delta$
 decreases substantially with decreasing temperature between $T_{a}$
 and $T_{b}$.  This loss of linewidth, already reported in ref.  3, is
 recovered almost discontinuously at $T_{b}$ and $\Delta(T)$ is
 approximately constant below that temperature.  This particular
 behavior is observed at Larmor frequencies of 67.00 and 45.75 MHz,
 but is absent in much lower magnetic fields.  This is emphasized in
 Fig.  2, where  $\Delta(T)$ of the central
 $^{27}$Al Zeeman transition for spectra measured at 67.00 and 26.00
 MHz, corresponding to applied magnetic fields of 6.4 and 2.5 T,
 respectively, is displayed.  While $\Delta(T)$ is approximately the
 same for both cases between $T_{a}$ and $T_{b}$, no discontinuous
 enhancement of $\Delta$ is manifest in the data set measured at 26
 MHz.

In Fig.  3.  we show three examples of $^{55}$Mn-NMR spectra, recorded
at a fixed frequency of 71.33 MHz and at temperatures of 0.48, 3.0 and
20 K. The central transition ($1/2 \leftrightarrow -1/2$) of the
$^{55}$Mn nuclei is centered at approximately 6.79 T. As the $^{27}$Al
nuclei, also the $^{55}$Mn nuclei carry a spin of $I = 5/2$ and thus
exhibit a relatively large electric-quadrupole moment.  Also here,
the quadrupolar wings are  expected to be widely spread out and
indeed, they cannot be resolved in our experiments.  Considering the
small lineshift, of the order of -0.3\%, as well as the moderate
linewidth monitored in the $^{55}$Mn-NMR central transition, we 
 conclude that the observed Mn-NMR signal originates from Mn nuclei
of non magnetic ions.  Because of the expected generation of large
hyperfine fields the resonance of the nuclei of magnetic Mn ions is
assumed to be outside  our observation range.  By comparing the
integrated intensities of the recorded central lines for $^{27}$Al and
$^{55}$Mn nuclei, and taking into account the differences in the
intrinsic NMR intensities for $^{27}$Al and $^{55}$Mn, we conclude
that at most a few percent of the Mn ions carry a magnetic moment.
\begin{figure}[t]
\begin{center}\leavevmode
\includegraphics[width=0.65\linewidth]{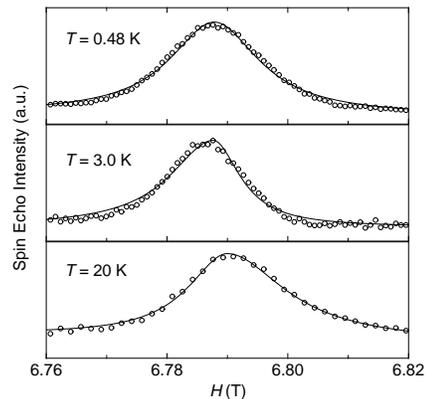}
\caption{
Central transition of the $^{55}$Mn-NMR spectra of 
Al$_{72.4}$Pd$_{20.5}$Mn$_{7.1}$ icosahedral quasicrystal measured at 
71.33 MHz and three temperatures.  The solid lines represent the
best fits to the data using a Lorentzian-type function with a degree
of asymmetry built into it. 
}
\protect
\label{Figure3}
\end{center}
\end{figure}
  As in the case of the $^{27}$Al NMR spectra, Fig.  3 reveals that
  the linewidth of the $^{55}$Mn central transition in the spectrum
  measured at 3.0 K is distinctly smaller than the corresponding
  linewidths at 0.48 and 20 K. Again upon decreasing $T$ an abrupt
  increase of the $FWHM$ $\Delta$ of the $^{55}$Mn signal is observed
  at $T_{b}$ in various fields of the order of 4 T or higher.  Various
  checks, with measurements performed using different conditions for
  the signal recording, indicate that this abrupt change of the
  linewidth is a rather robust feature for this quasicrystalline
  material and it is very unlikely that it is caused by extraneous
  artifacts of the measurements.

 We also investigated the NMR response of
 Al$_{72.4}$Pd$_{20.5}$Mn$_{7.1}$ by measurements of the temperature
 dependence of the spin-spin relaxation rate $T_{2}^{-1}$.  For this
 purpose the spin-echo lifetime $T_{2}^{\star}$ was first extracted
 from spin-echo decay curves, $i.e.$, curves of the echo-intensity as
 a function of the time delay $\tau$ between the two pulses of a
 $\pi/2 - \tau - \pi$ spin-echo sequence.  The effective rate
 $T_{2}^{-1}$ was then calculated via $T_{2}^{-1} = T_{2}^{\star - 1}
 - T_{1}^{-1}$.  The spin-lattice relaxation rate $T_{1}^{-1}$ was
 measured separately and was found to be much smaller than
 $T_{2}^{\star - 1}$ at all temperatures covered in our experiments. 
 The results for $T_{2}^{-1}(T)$, evaluated for the $^{27}$Al-NMR
 central transition at the Larmor frequency of 67.00 MHz, are
 displayed in Fig.  4.  We note the gradual increase of
 $T_{2}^{-1}$ with decreasing temperatures below $T_{a}$ and, again
 near $T_{b}$, its rather abrupt reduction by more than a factor of
 five.  This overall behavior is to be compared with previous
 observations involving metals with atoms containing unfilled
 3$d$-electron shells, such as Mn, where one often finds, with
 decreasing $T$, a progressive increase of the NMR linewidth, an
 increase of $(T_{1}T)^{-1}$ and, if any, a reduction of $T_{2}^{-1}$. 
 Both the reduction of the NMR linewidth and the increase of
 $T_{2}^{-1}$ with decreasing $T$ below $T_{a}$ are thus unexpected,
 but they seem to be related to each other.  Similarly, the sharp
 changes in the $FWHM$ and in $T_{2}^{-1}$ and to a lesser extent in
 $(T_{1}T)^{-1}$, observed at $T_{b}$, undoubtedly reflect a common
 cause.
\begin{figure}[t]
\begin{center}\leavevmode
\includegraphics[width=0.65\linewidth]{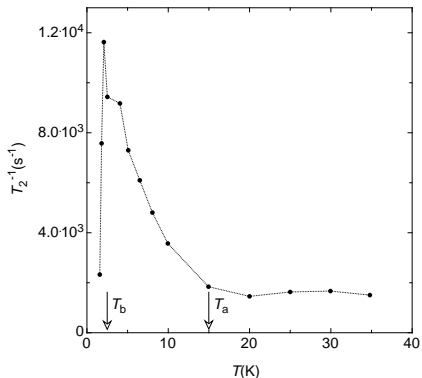}
\caption{ Spin-spin relaxation rate $T_{2}^{-1}$ as a function of
temperature measured at the $^{27}$Al-NMR central transition at 67
MHz.  Note the rapid increase of $T_{2}^{-1}$ with decreasing-$T$
below $T_{a}$ and the distinct maximum at $T_{b}$.  The dotted line is
a guide to the eye.  } \protect
\label{Figure4}
\end{center}
\end{figure}
In attempting a further characterization of the anomalous NMR response
of Al$_{72.4}$Pd$_{20.5}$Mn$_{7.1}$ at low temperatures, we have
measured the temperature dependence of the NMR spin-lattice relaxation
rate $T_{1}^{-1}$ under various experimental conditions.  Contrary to
what has been suggested previously\cite{Apih99} we found no evidence
for any appreciable contribution to $T_{1}^{-1}$ from quadrupolar
relaxation.  In Fig.  5 we display $(T_{1}T)^{-1}$ as a function of
$T$, with $T_{1}$ measured at the central transition of the $^{27}$Al
nuclei, for two different applied magnetic fields of 1.147 and 6.054
T, corresponding to Larmor frequencies of 12.70 and 67.00 MHz,
respectively.
\begin{figure}[t]
\begin{center}\leavevmode
\includegraphics[width=0.65\linewidth]{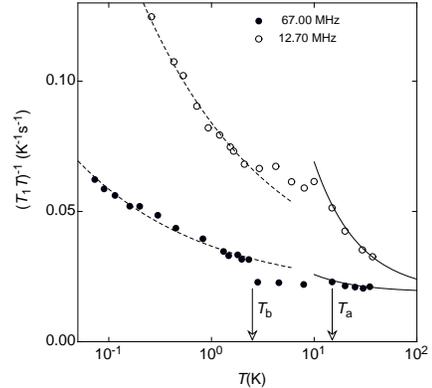}
\caption{ 
$(T_{1}T)^{-1}$ as a function of temperature measured at the
$^{27}$Al-NMR central transition at 67.00 and 12.70 MHz.  Above
$T_{a}= 15 $ K, $T_{1}^{-1}$ may be regarded as caused by the
conduction electrons and isolated Mn magnetic moments (solid lines). 
The latter contribution, however, turns into a power law below $T_{b}
= 2.5 $K (broken lines).  
} \protect
\label{Figure5}
\end{center}
\end{figure}
 Considering the relaxation data measured at 6.054 T, we note that
 below $T_{b}$, $(T_{1}T)^{-1}$ gradually increases with decreasing
 temperatures.  This is not unusual for quasicrystals.  For example, a
 qualitatively similar behavior, $i.e.$, an increase of $(T_{1}T)^{-1}$
 by more than an order of magnitude below approximately 20 K, has
 previously been reported for a nonmagnetic
 Al$_{70}$Re$_{8.6}$Pd$_{21.4}$ icosahedral
 quasicrystal\cite{Gavilano97} and very recently, a similar behavior
 has been observed in $T_{1}^{-1}(T)$ of
 Al$_{70}$Re$_{10}$Pd$_{20}$\cite{Rau2001}.  In the former case it has
 been suggested that this might reflect the critical nature of the
 itinerant-electron states whose existence in quasicrystals is
 presently a matter of debate\cite{Hafner99}.  More recently it has
 been suggested\cite{Dolinsek00} that paramagnetic impurities,
 inadvertently introduced during the synthesis of the material, may in
 general be the cause of the anomalous low temperature spin-lattice
 relaxation observed in quasicrystals.  However, this would lead to a
 strongly field-dependent spin-lattice relaxation rate
 $T_{1,imp}^{-1}$\cite{Benoit63,Giovannini71}, contrary to previous
 experimental observations\cite{Gavilano97} and therefore, this
 scenario cannot be regarded as being valid in all cases, indicating
 that the low-temperature behavior of $T_{1}^{-1}(T)$ of quasicrystals
 remains an interesting unsolved problem.

In our case, however, a small number of magnetic Mn ions has to be
considered.  The $\chi(T)$ data, exhibiting a Curie-Weiss-type feature
above 20 K, reveals a small paramagnetic Curie temperature and an
effective magnetic moment corresponding to a concentration $c \approx
1\%$ of magnetic Mn$^{3+}$ ions ($S = 2, L = 0$).  Consequently it
seems appropriate to consider the remaining magnetic Mn$^{3+}$ ions as
isolated magnetic moments, $i.e.$, ``paramagnetic impurities''.  For
this case the temperature and field dependences of the nuclear
relaxation rate $(T_{1}^{imp})^{-1}$, at temperatures where the random
fluctuations of the impurity moments are not correlated, is given
by\cite{Benoit63,Giovannini71}
\begin{equation}
	\frac{1}{T_{1}^{imp}T}  \propto c \cdot \frac{\langle S^{z} \rangle }{H} \cdot 
	\frac{\tau}{(1+(\omega \tau)^{2})} \; ,
	\label{eq:1}
\end{equation}
where $\tau$ is a time scale characterizing the spectrum of
fluctuations of the impurity (correlation time) and $\langle S^{z}
\rangle $ is the average $z-$component of the impurity spins induced
by $H$.  For simplicity we assume that $\langle S^{z} \rangle $
has a Curie-type temperature dependence\cite{Curie}.  The angular
Larmor frequency $\omega \propto H$.  Since our quasicrystalline
material is metallic, an additional Korringa-type contribution to the
spin-lattice relaxation rate via the conduction electrons,
$(T_{1}^{ce})^{-1}$, is expected\cite{KorringaDeviations}.  In total,
\begin{equation}
(T_{1}T)^{-1} = (T_{1}^{ce}T)^{-1} + (T_{1}^{imp}T)^{-1} .
	\label{eq:2}
\end{equation}

Indeed the experimental data above $T_{a}$ are fairly well represented
by invoking the above mentioned interpretation.  The solid lines in
Fig.  5 are based on Eq.  2, both with the same value of
$(T_{1}^{ce}T)^{-1} \approx 0.019$ K$^{-1}$s$^{-1}$, indicating that
the relaxation via conduction electrons is field-independent.  This
relaxation rate is very close to that reported in ref. 20 for
Al$_{75}$Pd$_{15}$Re$_{10}$ but by nearly a factor of 3 lower than the
corresponding rate observed in nonmagnetic
Al$_{70}$Pd$_{20}$Re$_{10}$\cite{Rau2001} and
Al$_{70}$Re$_{8.6}$Pd$_{21.4}$\cite{Gavilano97}.  This is rather
surprising because the electrical conductivities $\sigma (T)$ of the
Re alloys are, respectively, between one and two orders of magnitude
lower than $\sigma (T)$ of our Al-Mn-Pd
quasicrystal\cite{Dolinsek01,Bianchi97}. Thus this observation might
indicate that the electrical transport in quasicrystals is not simply
dictated by the number of itinerant charge carriers, but is dominated
by scattering processes that are not well understood.

Because of the factor $(1 + \omega^{2}\tau^{2})^{-1}$ in Eq.  (1), the
magnitude of the second term in Eq.  (2), due to the relaxation via
impurity spins, is influenced by the strength of the external field. 
Our fits, considering the respective field values and assuming a
temperature - and field - independent\cite{McHenry72} correlation time
$\tau$ in this temperature range, imply a very large value of $\tau =
7 \cdot 10^{-9}$ s$^{-1}$.  This is orders of magnitude larger than
those expected for dilute paramagnetic impurities in metals, such as
Gd impurities in LaAl$_{2}$\cite{McHenry72}.  Since there is no
obvious reason for an extremely weak coupling between the moments and
the conduction electrons, the correlation time $\tau$ may be large
because of the proximity of a magnetic phase transition at
temperatures of the order of $T_{a}$, slowing down the moment
fluctuations.

We note that the salient features in the $T$ dependence of
$T_{1}^{-1}$ are definitely field dependent below $T_{a}$.  The low
field data for $(T_{1}T)^{-1}$ exhibit a clear break in the slope
$\partial (T_{1}T)^{-1}/\partial T$ at or very near $T_{a}$ which may,
in the most straightforward way, be interpreted as a sudden variation
of $T_{1}$, due to a change in the interaction between fluctuating
impurities.

The high field data differs in the sense that, if at all, the break in
the slope of $(T_{1}T)^{-1}(T)$ at or very near $T_{a}$ is barely
visible, but instead a discontinuity in $(T_{1}T)^{-1}$ at $T_{b}$,
concomitant with the discontinuous line width enhancement, is
observed.  Both in high and low fields, the linewidths are nearly
temperature independent below $T_{b}$, but the relaxation rates
$T_{1}^{-1}$, vary as $(T_{1}T)^{-1} = (T_{1}^{ce}T)^{-1} + C(H) \cdot
T^{-0.35}$, represented by the broken lines in Fig.  5.  Here $C(H)
=$ 0.018 and 0.065 K$^{-0.65}$s$^{-1}$ for the frequencies of 67 and
12.7 MHz, respectively.  This assumes that the contribution of the
conduction electrons to the spin-lattice relaxation rate is not
altered very much at low temperatures, as suggested by the small
changes of $\rho(T)$ observed in our sample below 15
K\cite{Dolinsek01}.  In any case, the temperature and field
dependences of $\Delta$ and $T_{1}^{-1}$ below $T_{a}$ are very
unusual.  Although, our sample seems to be close to some kind of
magnetic instability above $T_{a}$, our observations for the low
temperature regime $T \leq 15$ K cannot be reconciled with
expectations, neither for non-magnetic hosts containing magnetic
impurities, nor for spin glasses or magnetically ordered systems.

In conclusion we report here the observation of unusual anomalies in
the NMR response of an icosahedral AlPdMn quasicrystal, which to our
knowledge have not been reported to occur in periodic crystals.  The
low-temperature features of Al$_{72.4}$Pd$_{20.5}$Mn$_{7.1}$ observed
in this work seem to indicate unusual variations in the stability of
transition metal (Mn) ionic moments in a non-periodic metallic matrix
at different temperatures.  At present we have no convincing
explanation for the observed phenomena (see also reference 3), but in
view of the covered temperature regimes, they must be related to
rather low characteristic energies.
 
This work was financially supported in part by the Schweizerische 
Nationalfonds (SNF).  We also acknowledge a special grant of the SNF 
for an inter-institutional collaboration between the Josef Stefan 
Institute, Ljubliana and ETH Z{\" u}rich.



\begin{thebibliography}{99}
\vspace*{-1cm}
\bibitem{Chernikov93} M.A. Chernikov, A. Bernasconi, C. Beeli and 
H.R. Ott, Europhys. Lett. {\bf 21} 767 (1993).

\bibitem{Simonet98} V. Simonet, F. Hippert, M. Audier, G. T. de 
Laissardi\`{e}re, Phys. Rev. B {\bf 58}, R8865, (1998).

\bibitem{Dolinsek01} J. Dolin${\rm \check s}$ek, M. Klanj${\rm \check s}$ek, T. Apih,
J.L. Gavilano, K. Gianno, H.R. Ott, J.M. Dubois and K. Urban, Phys. 
Rev.  B, accepted for publication.

\bibitem{Krajci98} M. Kraj${\rm \check c}$\'{i} and J. Hafner,
Phys.  Rev.  B {\bf 58}, 14110 (1998).

\bibitem{deLaissardiere00} G.T. de Laissardi\`{e}re and D. Mayou, Phys. Rev. 
Lett. {\bf 85}, 3273 (2000)

\bibitem{Bennett87} L.H. Bennett, M. Rubinstein, Xiao Gang and C.L. 
Chien, J. of Applied Phys. {\bf 61}, 4364 (1987).

\bibitem{Warren86} W.W. Warren, H.-S. Chen and G.P. Espinosa, Phys Rev. 
B {\bf 34}, 4902 (1986).

\bibitem{Shinohara93} T. Shinohara, Y. Yokoyama, M. Sato, A. Inoue, 
T. Matsumoto, J. of Phys.-Cond. matt. {\bf 5}, 3673 (1993)

\bibitem{Rodmar99} M. Rodmar, B. Grushko, N. Tamura, K. Urban and \"{O}. 
Rapp, Phys. Rev. B {\bf 60}, 7208, (1999).

\bibitem{Shastry94} A. Shastri, F. Borsa, D.R. Torgeson, J.E. Shield, 
and A.I. Goldman, Phys. Rev. B {\bf 50}, 15651 (1994).

\bibitem{Apih99} T. Apih, O. Plyushch, M Klanj${\rm \check s}$ek, and 
J. Dolin${\rm \check s}$ek, Phys.  Rev.  B {\bf 60}, 14695 (1999).

\bibitem{Gavilano97} J.L. Gavilano, B. Ambrosini, P. Vonlanthen, M. A. 
Chernikov, and H. R. Ott, Phys.  Rev.  Lett.  {\bf 79}, 3058 (1997).

\bibitem{Rau2001} D. Rau, J.L. Gavilano, Sh. Mushkolaj, P. Vonlanthen
and H. R. Ott, unpublished.

\bibitem{Hafner99} J. Hafner, Current-Opinion-in-Solid State and 
Materials Science, {\bf 4}, 289 (1999).

\bibitem{Dolinsek00} J. Dolin${\rm \check s}$ek, M Klanj${\rm \check 
s}$ek, T. Apih, A. Smontara, J.C. Lasjaunias, J.M. Dubois and S.J. 
Poon, Phys. Rev. B {\bf 62}, 8862 (2000).

\bibitem{Benoit63} H. Benoit, P. G. de Gennes and D. Silhoutte, 
Compt. Rend.  {\bf 256}, 3841 (1963).

\bibitem{Giovannini71} B. Giovannini, P. Pincus, G. Gladstone and A. 
J. Heeger, J. Phys. (Paris)  {\bf 32}, C1-163 (1971).

\bibitem{Curie} The conclusions are not significantly altered if a
Curie-Weiss type of temperature dependence for $\langle S_{z}\rangle$
is postulated.  One should also keep in mind that Eq. 
1 is already a high-temperature approximation, although, it works
surprisingly well even in some cases where the interactions among the
magnetic impurities are not really small.

\bibitem{KorringaDeviations} Deviations from a Korringa type
$T_{1}(T)$ have been observed at temperatures above 100 K for a
variety of quasicrystalline materials.  They have been attributed to
the existence of narrow features in the electronic density of states
at the Fermi level.  They need not be considered in our experiments.

\bibitem{Shinohara93a} T. Shinohara, A.P. Tsai, and
T. Matsumoto, Hyperfine Interactions {\bf 78}, 515 (1993).

\bibitem{Bianchi97} A.D. Bianchi, F. Bommeli, M. A. Chernikov, U. 
Gubler, L. Digiorgi and H. R. Ott, Phys. Rev. B  {\bf 55}, 5730 (1997).

\bibitem{McHenry72} M.R. McHenry, B.G. Silbernagel and J.H. Wernick, 
Phys. Rev. B {\bf 5}, 2958 (1972).

\end{thebibliography}
\end{document}